\documentclass{aa}
\usepackage{graphicx}
\usepackage{txfonts}
\usepackage{amstext}
\usepackage{color}

\begin{document}

   \title{Time lag in transient cosmic accreting sources}


   \author{G. S. Bisnovatyi-Kogan
          \inst{1}
          \and
          F. Giovannelli\inst{2}
          }
   \institute{Physics Department, Sapienza University of Rome, Piazzale Aldo Moro 5, 00185 Rome, Italy;\\
   National Research Nuclear University MEPhI (Moscow Engineering Physics Institute), Kashir\-skoe Shosse 31, Moscow 115409, Russia;\\
   Space Research Institute of Russian Academy of Sciences, Profsoyuznaya 84/32, Moscow 117997, Russia\\
              \email{gkogan@iki.rssi.ru}
         \and
             INAF - Istituto di Astrofisica e Planetologia Spaziali, Via del Fosso del Cavaliere, 100, 00133 Roma, Italy\\
             \email{franco.giovannelli@iaps.inaf.it}
             }

   \date{Received .....; accepted ........}


  \abstract
   { We develop models for time lag between the maxima of the source brightness in different wavelengths
during a transient flash of luminosity that is connected with a short-period increase of the mass flux onto
 the central compact object.}
   {We derive a simple formula for finding the time delay among events in different  wavelengths which is valid in general for all disk-accreting cosmic sources. We quantitatively also discuss a model for time-lag formation in active galactic nuclei (AGNs).}
   {In close binaries with accretion disks, the time lag is connected with  effects of viscosity that define a radial motion of matter
   in the accretion disk. In AGN flashes, the infalling matter has a low angular momentum, and the time lag is defined by the free-fall time to the gravitating center.  }
   {We show the validity of these models by means of several examples of galactic and extragalactic accreting sources.}
   {}


   \maketitle
%

\section{Introduction}
The optical behavior of the Be star in the high-mass X-ray transient A0535+26/ HDE245770, as discussed by
Giovannelli \& Sabau-Graziati (\citeyear{2011AcPol..51b..21G}), shows that the luminosity at periastron
is typically enhanced by $\sim 0.02$ to a few tenths magnitude
, and the X-ray outburst occurs eight days after the periastron.
Indeed, an increase of the mass flux occurs
at periastron. This flush reaches the external part of the
temporary accretion disk around the neutron star and moves to the hot central parts of the accretion disk and the
 neutron star surface. The time necessary for this way is dependent on the turbulent viscosity in the accretion
 disk, as discussed  by  Giovannelli, Bisnovatyi-Kogan \& Klepnev (\citeyear{2013A&A...560A...1G}) (GBK13).
Giovannelli et al. (\citeyear{2015AcA....65..107G}) discussed the behaviour of A0535+26/HDE245770 during 2014 by using the GBK13 ephemeris -- JD$_{\rm opt-outb}$ = JD$_0$(2,444,944) $\pm$ n(111.0 $\pm$ 0.4) days,
in order to describe the passages of the neutron star at the periastron  at the 106th and 108th cycles after the
optical event occurred on December 5, 1981 (811205-E), which suggested that the incoming X-ray outburst occurred on
December 13, 1981 (Nagase et al. \citeyear{1982ApJ...263..814N}). The results of Giovannelli et al. (\citeyear{2015AcA....65..107G})
 and the derived relationship $\Delta$V$_{\rm mag}$ vs I$_x$ can predict not only the arrival time of X-ray outbursts, but also its intensity I$_x$.

GBK13 have constructed a quantitative model for explaining the time delay between optical and X-ray outbursts.
The mechanism proposed by GBK13 to explain the X-ray -- optical delay in A 0535+26/HDE 245770 is based on an enhanced mass flux propagation through the viscous accretion disk.
The observed time delay is related to the motion of a high-mass flux region from the outer boundary of the neutron star Roche lobe to the Alfv\'{e}n surface that is due to the action of the $\alpha$--viscosity.
This mechanism, known as the UV - optical delay (the delay of the extreme ultraviolet (EUV) flash with respect to the optical flash), was observed and modeled for cataclysmic variables (e.g., Smak \citeyear{1984AcA....34..161S},\citeyear{1998AcA....48..677S}; Lasota \citeyear{2001NewAR..45..449L}).

Lags between the optical and X-ray maxima in flashes had also
been observed in active galactic nuclei (AGNs
).
Nandra et al. (\citeyear{1998ApJ...505..594N}) found a delay of $\sim\text{4}$  days between UV and X-ray emissions in NGC 7469; Maoz, Edelson \& Nandra (\citeyear{2000AJ....119..119M})  found a delay of $\sim$ 100 days between optical and X-ray emissions in the Seyfert galaxy NGC 3516; Marshall, Ryle \& Miller (\citeyear{2008ApJ...677..880M})  found a delay of $\sim$ 15 days between optical and X-ray emissions in Mkr 509, and Doroshenko et al. (\citeyear{2009AstL...35..361D}) found a delay of $\approx$ 10 days between R, I and X-ray luminosities in the Seyfert galaxy 3C 120; Shemmer et al. (\citeyear{2003MNRAS.343.1341S}) found a delay of $2.4 \pm 1.0$ days between optical and X-ray emissions in NGC 4051.
The time lags observed in AGNs are much shorter than we would expect from the scaling in formulae (7) and (19).
In the disk accretion model we expect time lags
orders of magnitude larger than in galactic X-ray transients, and judging from observations, these lags are of the same order.
We suggest that this is connected with a much faster contraction of accreting matter, which occurs in the case of quasi-spherical accretion of matter with low angular momentum.

In this paper we derive a simple formula for finding the time delay among events in different wavelengths, which is valid in general for all disk-accreting cosmic sources, and
quantitatively discuss a model for time-lag formation in AGNs.

\section{Equations of the accretion disk structure}

When the characteristic time of variability of the mass flux along the accretion disk is longer than the
relaxation time of the local disk equilibrium, it is possible to use the approximation of local equilibrium
Shakura (\citeyear{1972AZh....49..921S}), see also Bisnovatyi-Kogan (\citeyear{2011spse.book.....B}), to calculate the transient disk structure.
The equilibrium along a radius of the accretion disk around a star with a mass $M$
 is determined by the Keplerian rotational velocity $\Omega_K$

\begin{equation}
\label{omega}
\Omega=\Omega_K=\left(\frac{GM}{r^3}\right)^{1/2}.
\end{equation}
Writing the equation of the vertical equilibrium in approximate algebraic form, we obtain

\begin{equation}
\label{h}
 h=\sqrt 2 \frac{v_s}{\Omega},
\end{equation}
where $v_s=\sqrt{P/\rho}$ is the speed proportional to the sound velocity, $P$ and $\rho$ are the (gas + radiation) pressure and
density at the symmetry plane of the accretion disk, and $h$ is the semi-thickness of the accretion disk.
 The specific angular momentum $l$ of the matter in the accretion
disk is connected to the rotation velocity as

\begin{equation}
\label{l}
l=r\,v_\phi=r^2\Omega.\end{equation}
The mass flux through the disk at radius $r$ is connected to the radial velocity $v_r$ as

\begin{equation}
\label{mflux}
\dot M=-4\pi h\rho r v_r, \quad \dot M>0,\quad v_r<0.
\end{equation}
We use an $\alpha$ approximation for the turbulent viscosity (Shakura, 1972) when the $(r\phi)$ component of the stress tensor
$t_{r\phi}$ is written as

\begin{equation}
t_{r\phi}=\alpha\, P,
\label{trphi}
\end{equation}
where the phenomenological  non-dimensional parameter $\alpha \le 1$. The condition of
stationarity of the angular momentum, in which the outward viscous radial flux of the angular momentum is
balanced by the angular momentum of the inward flux of the mass, is written as (see, e.g., Bisnovatyi-Kogan, 2011)

\begin{equation}
r^2 h \alpha P=\frac{\dot M}{4\pi}(l-l_{in})
\label{angmom}
.\end{equation}
The main input into the time lag comes from the outer regions of the disk with $l\gg l_{in}$. Then we have from Eqs.
(\ref{mflux}) and (\ref{angmom}) the expression for the radial velocity in the form

\begin{equation}
v_r=-\alpha\frac{v_s^2}{v_\phi}.
\label{vr}
\end{equation}
We also define the surface density $\Sigma$, and write Eq. (\ref{angmom})
in light of
 Eq.(\ref{l}), using condition $l\gg l_{in}$, in the form

\begin{equation}
\Sigma=2\rho h,\quad \dot M \Omega=4\pi \alpha P h.
\label{sigma}
\end{equation}
The equation of the local thermal balance in the accretion disk, when the heat produced by viscosity $Q_+$ is entirely emitted
through the sites of the optically thick accretion disk with a total flux $Q_-$, at  $l\gg l_{in}$ is written as (see, e.g., Bisnovatyi-Kogan, 2011)

\begin{equation}
\frac{3}{2}\dot M \Omega^2=\frac{16\pi ac T^4}{3\varkappa \Sigma}.
\label{theq}
\end{equation}
Here $T$ is the temperature in the symmetry plane of the accretion disk, $a$ is the constant of the radiation energy density,
$c$ is the speed of light, and $\varkappa$ is the Thompson (scattering)  opacity of the matter.

\section{Calculation of the time lag in the disk accretion model for galactic X-ray sources}

 Based on Eqs. (\ref{sigma}) and (\ref{h}), Eq. (\ref{theq})
may be written as

\begin{equation}
\frac{T^4}{\rho v_s}=\frac{9}{8\sqrt 2}\frac{\dot M \varkappa\Omega}{\pi ac}.
\label{theq1}
\end{equation}
From the second relation in Eq. (\ref{sigma}), using
Eq. (\ref{h}), we obtain the relation

\begin{equation}
\rho v_s^3=\frac{\dot M}{4\pi\sqrt 2}\frac{v_\phi^2}{\alpha r^2}.
\label{vs}
\end{equation}
The time lag in transient X ray sources is typically formed in the regions of the
accretion disk that are dominated by gas pressure  with $P=P_g$,
and the scattering opacity $\varkappa=0.2(1+X)$, where $X$ is the hydrogen mass fraction.
The  gas constant $\cal R$  for the hydrogen-helium fully ionized plasma and the gas pressure are
written as (Landau and Lifshitz, \citeyear{1980stph.book.....L})

\begin{equation}
 P_g=\rho{\cal R} T, \quad {\cal R} = \frac{k}{\mu m_p},\quad \mu=\frac{4}{5X+3}.
\label{pgas}
\end{equation}
Here $k$ is the Boltzmann constant and $m_p$ is the proton mass.
By multiplying Eqs. (\ref{theq1}) and (\ref{vs}), we obtain, based on Eq. (\ref{pgas}), the relation

\begin{equation}
T^5=\frac{9}{64\pi^2}\frac{\dot M^2 \varkappa \Omega^3}{\alpha ac \cal R}.
\label{t5}
\end{equation}
Our main goal is to express the local parameters of the accretion disk through the observed parameters:
mass of the compact star (or a black hole) $M$, mass flux into the star $\dot M$ measured through its bolometric
luminosity, and the effective temperature $T_{eff}$ of the object, evaluated from its spectra. The radius where radiation is the
most effective during the transient accretion may be found from the thermal equilibrium condition by accepting that
the radiation is emitted from the sides of the disk with an effective temperature $T_{eff}$

\begin{equation}
Q_+=\frac{3}{8\pi} \dot M \frac{GM}{r^3}=\sigma T_{eff}^4,
\label{te1}
\end{equation}
giving

\begin{equation}
r=\left(\frac{3}{8\pi}\frac{GM\dot M}{\sigma T^4_{eff}}\right)^{1/3}.
\label{te2}
\end{equation}
Using Eqs. (\ref{omega}) and (\ref{te2}), Eq. (\ref{t5}) is written in the form

\begin{equation}
T^5=\left(\frac{3\dot M}{8\pi }\right)^{1/2}\frac{\varkappa}{ac\alpha  \cal R}(\sigma T^4_{eff})^{3/2}.
\label{t51}
\end{equation}
The radial velocity from Eq. (\ref{vr}) may be written as

\begin{equation}
v_r=-\alpha{\cal R} T \sqrt{\frac{r}{GM}}.
\label{vr1}
\end{equation}
Based on Eqs. (\ref{te2}) and (\ref{t51}), we obtain

\begin{equation}
v_r=-\left(\frac{3}{8\pi}\right)^{4/15}\left(\frac{\varkappa}{ac}\right)^{1/5}
\frac{(\alpha {\cal R})^{4/5}}{(GM)^{1/3}}\dot M^{4/15} (\sigma T_{eff}^4)^{2/15}.
\label{vr2}
\end{equation}
For a given $\dot M$ in the flash, the value of $T_{eff}$ changes in time, while
the wave of a large $\dot M$ moves to the star along the accretion disk (GBK13; Giovannelli et al. 2015; Bisnovatyi-Kogan et al. \citeyear{2014IJMPS..2860186B}).
The equation  for the radius of the hot layer flash-wave moving to the star is written as

 \begin{equation}
\frac{dr}{dt}=v_r.
\label{rt}
\end{equation}
We can reduce it to the variable $T_{eff}$ instead of $r$, using Eqs. (\ref{te2}) and (\ref{vr2}), when it is written in the form

 \begin{eqnarray}
\frac{dT_{eff}}{dt}=-\frac{3}{4}\frac{v_r}{r}T_{eff} \qquad\qquad\nonumber\\
=\frac{3}{4}\left(\frac{8\pi}{3}\right)^{1/15}\left(\frac{\varkappa}{ac}\right)^{1/5}\frac{(\alpha {\cal R})^{4/5}}{(GM)^{2/3}}
    \frac{T_{eff}}{\dot M^{1/15}}(\sigma T_{eff}^4)^{7/15}.
\label{rt1}
\end{eqnarray}
Writing Eq. (\ref{rt1}) as

 \begin{equation}
\frac{dT_{eff}}{dt}=A\,T_{eff}^{43/15}, \quad A=\frac{3}{4}\left(\frac{8\pi}{3}\right)^{1/15}\left(\frac{\varkappa}{ac}\right)^{1/5}\frac{(\alpha {\cal R})^{4/5}}{(GM)^{2/3}}
    \frac{\sigma^{7/15}}{\dot M^{1/15}},
\label{rt2}
\end{equation}
we obtain its solution in the form

\begin{equation}
-\frac{15}{28\,A}\,T_{eff}^{-28/15}=t+const.
\label{rt3}
\end{equation}
For the initial condition $T_{eff}(0)=T_0$, we find the time $\tau$, when the effective temperature $T_{eff}=T_1$, by the relation

 \begin{equation}
\tau=\frac{15}{28\,A}(T_0^{-28/15}-T_1^{-28/15}).
\label{rt4}
\end{equation}
If $T_0$ corresponds to the maximum of the optics and $T_1$ corresponds to the X-ray maximum, then the value of $\tau$
represents the time lag between these two maxima, which for the transient X-ray source \\
A0535+26/HDE245770 is
close to eight days (Giovannelli \& Sabau-Graziati, 2011). For $T_1\gg T_0$, using Eq. (\ref{rt2}), we have

\begin{equation}
\tau=\frac{15}{28\,A}\,T_0^{-28/15}=\frac{5}{7}\left(\frac{3}{8\pi}\right)^{1/15}\left(\frac{ac}{\varkappa}\right)^{1/5}\frac{(GM)^{2/3}}{(\alpha {\cal R})^{4/5}}
    \frac{\dot M^{1/15}}{(\sigma T_0^4)^{7/15}}
\label{rt5}
.\end{equation}
By substituting numbers into Eq. (\ref{rt5}) and introducing dimensionless values

 \begin{equation}
m=\frac{M}{M_\odot},\,\,\, \dot m=\frac{\dot M}{10^{-8}M_\odot/{\rm year}},\,\,\, T_4=\frac{T_0}{10^4 {\rm K}},
\label{rt6}
\end{equation}
we obtain the value of the time lag between optical and X-ray maxima in the transient Be/X-ray source, resulting
from a rapid increase in the mass flux of matter with $X=0.7$ through the accretion disk, using Eq. (\ref{pgas}), in the form

 \begin{equation}
\tau=6.9\frac{m^{2/3}\dot m^{1/15}}{\alpha^{4/5} T_4^{28/15}}\,\,{\rm days}.
\label{rt7}
\end{equation}
This formula is useful to from an approximate idea about the time lag between optical and X-ray maximum emissions in a cosmic disk-accreting source, regardless of whether it is a white dwarf, a neutron star, or a black hole. The time lag depends
 on the mass as ($\sim m^{2/3}$), and only very weakly on the mass flux, $\dot m^{1/15}$. The dependence on viscosity $\alpha$ and on $T_4$ is stronger.

For {\bf A0535+26/HDE245770}, using the viscosity $\alpha = 0.1$, as derived by the GBK13 model, $m = 1.4$ and $\dot{m} = 1$, we obtain from Eq. (26) $\tau=54/T_4^{28/15}$ days. Taking into account the experimental time lag $\tau_{\rm exp} \simeq 8$ days, the derived temperature is $T_4=2.78$.
Owing to the very weak dependence of $\tau$ on the accretion rate, the time lag in this source remains practically the same for flashes of different intensity, like the flash observed by de Martino et al. (\citeyear{1989ESASP.296..519D}), with a mass accretion rate $\dot{M}\simeq 7.7 \times 10^{-7}$ M$_\odot$ yr$^{-1}$, $\dot m = 77$.
When we use the neutron star mass M = 1.5 M$_\odot$ (Giovannelli et al. \citeyear{2007A&A...475..651G}), the viscosity coefficient  $\alpha$ = 0.15, and T$_4 \simeq 2.8$ (de Loore et al.  \citeyear {1984A&A...141..279D}), the time delay computed with Eq. (26), $\tau=8.06$ days, coincides with the experimental time delay (X-ray - optical) of $\sim 8$ days.

 When we assume the mass of a white dwarf $M = 0.97$ M$_\odot$ (Giovannelli et al. \citeyear{1983AcA....33..319G}), a mass accretion rate $\dot{M} = 4 \times 10^{17}$ g s$^{-1} \approx 6.3\times 10^{-9}$ M$_\odot yr^{-1}$, $\dot m=0.63$ (Giovannelli \& Sabau-Graziati \citeyear{2012AcPol..52a..11G}), a viscosity $\alpha$ = 0.2 (e.g. Smak \citeyear{1998AcA....48..677S}), and $T_4 = 4$ (Hameury et al. \citeyear{1999MNRAS.303...39H}), the time delay in the cataclysmic variable {\bf SS Cygni} computed with Eq. (26) is $\tau \simeq 1.8$ days. The experimental time delay (UV -- optical) is 0.9--1.4 days (Wheatley, Mauche \& Mattei \citeyear{2003MNRAS.345...49W}), and for $\alpha=0.3,$ we obtain
$\tau=1.35$ days, which is well inside the error box.

When we assume the mass of a neutron star $M = 1.4$ M$_\odot$ (Waterhouse et al. \citeyear{2016MNRAS.456.4001W}), a mass accretion rate $\dot{M} = 4 \times 10^{17}$ g s$^{-1} \approx 6.3\times 10^{-9}$ M$_\odot yr^{-1}$, $\dot m=0.63$ (Yamaoka et al. \citeyear{2011ATel.3686....1Y}; Meshcheryakov et al. \citeyear{2013ATel.5114....1M}), a viscosity $\alpha$ = 0.2, and T$_4 \simeq 2.8$, the time delay in the X-ray source {\bf Aql X-1} computed with Eq. (26) is $\tau \simeq 4.4$ days. The observed time delay (X-ray -- optical) of $\sim 3$ days (Shahbaz et al. \citeyear{1998MNRAS.300.1035S}) is better reproduced for $\alpha=0.3$, when $\tau=3.2$ days follows from Eq. (26).

 When we
assume the mass of a black hole $M \simeq 7$ M$_\odot$ (Orosz \& Baylin \citeyear{1997ApJ...477..876O}), and the approximate relationship $M/\dot{M} \sim 10^7$ yr (Karaku{\l}a, Tkaczyk \& Giovannelli \citeyear{1984AdSpR...3..335K})  in the black hole X-ray transient {\bf GRO J1655-40}, we have $\dot{M} \sim 7 \times 10^{-7}$ M$_\odot yr^{-1}$. For  $T_4 = 3$, we obtain from  Eq. (26) a time delay
$\tau \simeq 15.6$ days at $\alpha$ = 0.2, and $\tau \simeq 11.3$ days at $\alpha$ = 0.3. The  optical precursor was observed $\sim 6$ days before the X-ray flash (Orosz et al. \citeyear{1997ApJ...478L..83O}), therefore Eq. (26) overestimates the observed value. This time delay was observed in the X-ray flash of the LMXB X-ray nova with a black hole, which is typically  connected with instability
that develops in the accretion
disk after a sufficiently high mass is accumulated
in the quiescent state. When an instability develops, it leads to non-stationary behavior with higher viscosity. This explains why Eq. (26), which has been obtained for a stationary accretion disk, overestimates the value of the time lag. However, when we use a higher viscosity value, for example, $\alpha = 0.6$, the time lag computed with Eq. (26) is $\tau \simeq 6.5$ days, in agreement with the observed time delay.

\section{Time lags in AGN observations}

Supermassive black holes in AGNs radiate as a result of accretion, and time lags similar to galactic X-ray sources are observed during their sporadic variability, see Marshall, Ryle \& Miller (\citeyear{2008ApJ...677..880M}), Nandra et al.( \citeyear{1998ApJ...505..594N}), and other references below. The region of the optical emission in AGNs is situated somewhat closer to the black hole than in galactic sources, and the choice between radiative and gas pressure is not evident. To make this choice, we compared local solutions for the accretion disk structure and determined the boundary between the radiation-supported and gas-supported regions. The solution of the accretion disk structure in a local approximation gives the  expressions for the pressure (in CGS units) in the equatorial plane for radiation-dominated ($P_r$), and gas-dominated ($P_g$) regions as (Bisnovatyi-Kogan, 2011)

\begin{eqnarray}
\label{prg}
P_r=\frac{10^{16}}{\alpha\,m\,x^{3/2}},\qquad\qquad \nonumber \\
\qquad\qquad P_g=\rho{\cal R}T=1.39\cdot 10^{18}\frac{{\dot m}^{4/5}}{(\alpha m)^{9/10}\, x^{51/20}}
.\end{eqnarray}
At the boundary between two regions these pressures are equal, which determines the value of $x_b$ as

\begin{equation}
\label{prg1}
x_b=110\,(\alpha m)^{2/21} {\dot m}^{16/21}.
\end{equation}
Here and below, we use the notations

\begin{equation}
\label{prg2}
x=\frac{r\, c^2}{GM}, \quad m=\frac{M}{M_\odot},\quad \dot m =\frac{\dot M\, c^2}{L_c}, \text{ and}\quad L_c=\frac{4\pi cGM}{\kappa}.
\end{equation}
The disk effective temperature at the boundary $x=x_b$ follows from Eqw. (\ref{te1}), (\ref{prg1}), and (\ref{prg2}) as

\begin{equation}
\label{prg3}
T_{eff}=\left(\frac{3}{8\pi}\frac{GM\dot M}{\sigma r_b^3}\right)^{1/4}=\left(\frac{3}{2}\frac{c^5}{\varkappa\sigma GM_\odot}\frac{\dot m}{mx^3}\right)^{1/4}
=\frac{1.8\cdot 10^6\,\, K}{\alpha^{1/14}(m \dot m)^{9/28}}.
\end{equation}
For the characteristic temperature of the optical emission $T_4=2.5$, we obtain from Eq. (\ref{prg3}) the relation at which this temperature is situated
at the border between gas- and radiation-dominated regions as

\begin{equation}
\label{prg4}
\frac{1.8\cdot 10^6\,\, K}{\alpha^{1/14}(m \dot m)^{9/28}}=2.5\cdot 10^4\text{ and}\quad m\dot m=\frac{6\cdot 10^5}{\alpha^{2/9}}=8.6\cdot 10^5\end{equation}
for $\alpha=0.2$. The accretion rate at the critical Eddington luminosity corresponds to $\dot m\approx 16$ for a Schwarzschild BH and to $\dot m \approx 2.5$ for the
limiting Kerr metric. In most AGNs, the optical emission therefore
comes from the radiation-dominated regions, in contrast to the accretion into a binary galactic BH.

The formula for the time lag for the optics coming from a radiation-dominated disk region may be obtained in the same way as for the gas-dominated regions.
Now

\begin{equation}
\label{pr1}
P=P_r=\frac{aT^4}{3},\quad v_s^2=\frac{P_r}{\rho}=\frac{aT^4}{3\rho}.
\end{equation}
By multiplying Eqs. (\ref{theq1}) and (\ref{vs}), we obtain the relation

\begin{equation}
\label{pr2}
aT^4 v_s^2=\frac{9}{64\pi^2}\frac{\dot M^2\varkappa\Omega^3}{\alpha c}.
\end{equation}
By dividing Eqs. (\ref{theq1}) and (\ref{vs}), we obtain

\begin{equation}
\label{pr3}
\frac{aT^4}{\rho^2 v_s^4}=\frac{9\alpha}{2}\frac{\varkappa}{c \Omega}.
\end{equation}
Using Eq. (\ref{pr1}), we obtain

\begin{equation}
\label{pr4}
aT^4=\frac{2c\Omega}{\alpha\varkappa}.
\end{equation}
When we use Eq. (\ref{pr4}) in Eq. (\ref{pr2}), we obtain

\begin{equation}
\label{pr5}
v_s^2=\frac{9}{128\pi^2}\frac{({\dot M}\varkappa\Omega)^2}{c^2}.
\end{equation}
When we use Eqs. (\ref{omega}), (\ref{te2}), and (\ref{pr5}) in Eq. (\ref{vr}), we obtain  the radial velocity
in the accetion disk as a function of the effective temperature in the form

\begin{equation}
\label{pr6}
v_r=-\frac{3\alpha}{16\pi}\left(\frac{3}{8\pi}\right)^{1/6}
\frac{\varkappa^2{\dot M}^{7/6}(\sigma T_{eff}^4)^{5/6}}{c^2(GM)^{1/3}}.
\end{equation}
Similar to the gas-dominated case above, we obtain the equation for $T_{eff}$ as

\begin{equation}
\label{pr7}
\frac{d T_{eff}}{dt}=\frac{9\alpha}{64}\left(\frac{8\pi}{3}\right)^{1/6}
\frac{\varkappa^2{\dot M}^{5/6}(\sigma T_{eff}^4)^{7/6}}{c^2(GM)^{2/3}}T_{eff}.
\end{equation}
Writing this equation as

\begin{equation}
\label{pr8}
\frac{d T_{eff}}{dt}=BT_{eff}^{17/3},\quad
B=\frac{9\alpha}{64\pi}\left(\frac{8\pi}{3}\right)^{1/6}
\frac{\varkappa^2{\dot M}^{5/6}\sigma^{7/6}}{c^2(GM)^{2/3}},
\end{equation}
we obtain the expression for the time lag $\tau_r$ between the effective temperature $T_{eff}=T_0$, and
$T_{eff}=T_1 \gg T_0$, in the radiation-dominated disk region
as

\begin{equation}
\label{pr9}
\tau_r=\frac{896\pi}{27\alpha}\left(\frac{3}{8\pi}\right)^{1/6}
\frac{c^2(GM)^{2/3}}{\varkappa^2{\dot M}^{5/6}(\sigma T_0^4)^{7/6}}.
\end{equation}
Using notations from Eq. (\ref{rt6}), we obtain the expression for the time lag $\tau_r$ in the form

\begin{equation}
\label{pr10}
\tau_r=4.9\times 10^7
\frac{m^{2/3}}{\alpha{\dot m}^{5/6} T_4^{14/3}}\,\, {\rm days}.
\end{equation}

For AGNs we have used the value of $T_4=2.5$
when no direct data were available. For the $\alpha$-viscosity we have used a  value $\alpha$ =0.2 when no other indications were available.

The application of Eq. (\ref{pr10}) to some AGNs gives results that are much longer than those observed. This means that the mechanism responsible for the time lag in AGNs is different from that of galactic transient accreting sources. In particular, we have the estimations of the time lag of the objects listed below.

 In {\bf Mrk 509}  the mass of the supermassive black hole (SMBH) is estimated as $M \simeq 1.4 \times 10^8$ M$_\odot$ (Peterson et al. \citeyear{2004ApJ...613..682P}), and the mass flux $\dot{M} \sim 14$ M$_\odot yr^{-1}$ (Karaku{\l}a, Tkaczyk \& Giovannelli \citeyear{1984AdSpR...3..335K}). For the viscosity coefficient $\alpha$ = 0.2 and $T_4 = 2.5,$ the time delay computed with the formula (\ref{pr10}) is $\tau \simeq 2.2\times 10^4$ days, while the observational delay is $\tau_{obs}\simeq 15$ days (Marshall, Ryle \& Miller \citeyear{2008ApJ...677..880M}). Mehdipour et al. (\citeyear{2011A&A...534A..39M}) observed no significant time lag between optical and X-ray variability. In our model of the time lag in AGNs, which is presented in the next section, the time lag with optics preceding X-rays is connected with the
tidal disruption of stars approaching an SMBH. The type of the flare and a real time lag strongly depend on the properties of the disrupted star and may be very different. Therefore the absence of any time lag found by Mehdipour et al. (\citeyear{2011A&A...534A..39M}) may indicate a different (probably more compact) disrupted star, or any other physical mechanism that might have caused the flare.

 In {\bf NGC 7469}  the mass of SMBH is estimated as $M \simeq 10^7$ M$_\odot$ (Peterson et al. \citeyear{2004ApJ...613..682P}; Onken et al. \citeyear{2004ApJ...615..645O}), and the mass flux $\dot{M} \sim 1$ M$_\odot yr^{-1}$ (Karaku{\l}a, Tkaczyk \& Giovannelli \citeyear{1984AdSpR...3..335K}). For the viscosity coefficient $\alpha$ = 0.2, and $T_4 = 2.5$ the time delay computed with Eq. (\ref{pr10}) is $\tau \simeq 3.4\times 10^4$ days, while the observational time delay (UV -- optical) is $\tau_{obs}\simeq 4$ days (Nandra et al. \citeyear{1998ApJ...505..594N}).

 In {\bf 3C 120} the mass of SMBH is estimated as  $M \simeq 5.5 \times 10^7$ M$_\odot$ (Onken et al. \citeyear{2004ApJ...615..645O}; Vestergaard \& Peterson \citeyear{2006ApJ...641..689V}), and
the mass flux $\dot{M} \sim 5.5$ M$_\odot yr^{-1}$ (Karaku{\l}a, Tkaczyk \& Giovannelli \citeyear{1984AdSpR...3..335K}). For the viscosity coefficient $\alpha$ = 0.2 and $T_4 = 2.5,$  the time
delay computed with Eq. (\ref{pr10}) is $\tau \simeq 2.5\times 10^4$ days, while the observational time delays in this object are much shorter.

 Different time lags (X-ray--R,I,B) in 3C 120 have been studied by  Doroshenko et al. (\citeyear{2009AstL...35..361D}). The results of this paper are
rather controversial, and cannot be explained by a unique model. The authors report
  that the Rc- and Ic-band flux variations
lag significantly behind the B-band fluxes by 3.9
and 6.2 days, respectively, when the position of the CCF
centroid is considered at 0.8r$_{\rm max}$. In addition, the
X-ray variability on a long timescale also lags behind
the B-band variations by 5.3 days. However,
the confidence level of this estimate is only 87\%. A
more detailed analysis of the correlation between the
X-ray and optical emissions revealed a fairly complex
picture: the degree of correlation between the optical
and X-ray flux variations is different at different times. They also claim in the discussion
 that according to their data,
the X-ray emission in 3C 120 lags behind rather than
leads the optical emission when they take the
strong CCF asymmetry into account. This fact is also difficult to
reconcile with the simple reprocessing model.
We explain this particular type of  time lag  by the model described in the next section, where
we use for our estimations  $\tau_{lag}\simeq 10$ days. Other types of the time lag mentioned
in this paper could be explained by a reprocessing model or any other model.
Doroshenko et al.(\citeyear{2009AstL...35..361D}) conclude:
{\it "According to our data, the X-ray events may
sometimes lead the optical ones (episode 3), sometimes
occur almost simultaneously (episode 1) or
may lag behind."} This clearly shows that the results of this paper cannot be interpreted in only one way.

  In {\bf NGC 3516} the mass of SMBH is estimated as  $M \simeq 3.17 \times 10^7$ M$_\odot$  (Denney et al. \citeyear{2010ApJ...721..715D}), and the mass flux $\dot{M} \sim 3.17$ M$_\odot yr^{-1}$ (Karaku{\l}a, Tkaczyk \& Giovannelli \citeyear{1984AdSpR...3..335K}). For the viscosity coefficient $\alpha$ = 0.2 and $T_4 = 2.5,$  the time
delay computed with Eq. (\ref{pr10}) is $\tau \simeq 2.8\times 10^4$ days, while the experimental time delay (X-ray--optical) is  $\tau_{obs}\simeq 100$ days (Maoz, Edelson \& Nandra \citeyear{2000AJ....119..119M}).

 In {\bf NGC 4051} the mass of SMBH is estimated as  $M \simeq 5 \times 10^5$ M$_\odot$  (Shemmer et al. \citeyear{2003MNRAS.343.1341S}), and the mass flux $\dot{M} \sim 0.05$ M$_\odot yr^{-1}$ (Karaku{\l}a, Tkaczyk \& Giovannelli  \citeyear{1984AdSpR...3..335K}). For the viscosity coefficient $\alpha$ = 0.2 and $T_4 = 2.5,$  the time delay computed with Eq. (\ref{pr10}) is $\tau \simeq 5.6\times 10^4$ days, while the experimental time delay (X-ray--optical) is $\tau_{obs}= 2.4 \pm 1.0$ days (Shemmer et al. \citeyear{2003MNRAS.343.1341S}). In later observations, Alston et al. (\citeyear{2013MNRAS.429...75A}) did not find
  any correlation between  optical and X-ray light curves. This behavior is similar to the observational properties of MRK 509
that we discussed above, and may be explained in the same way.

The tidal disruption of a star by an SMBH was discovered by ASAS-SN with robotic telescopes in the center of the galaxy PGC 043234 and was followed by optical and X-ray instruments, including Swift and XMM (Miller et al. \citeyear{2015Natur.526..542M}). The discovery magnitude indicated a substantial flux increase over archival optical images of this galaxy. Archival X-ray studies rule out the possibility that PGC 043234 has a standard active nucleus that could produce bright flaring, because it was not detected in the ROSAT All-sky survey
(Voges, Aschenbach, Boller, et al. \citeyear{1999A&A...349..389V}). As was noted, the multiwavelength light curves in Fig. 1 of Miller et al. (\citeyear{2015Natur.526..542M}) clearly indicate a tidal disruption event. The authors noted that the X-ray points in this figure carry relatively large errors, but it is distinctly visible in this figure that
the maximum of the X-ray flux occurs about five days later than the maximum in the optical band. This means that the lime lag in the outburst in the nucleus of the galaxy PGC 043234 has an evident similarity with the outbursts in the five nuclei described above. The timescale of the lag in this nucleus is also short and cannot be explained by the processes
within a radiation-dominated region in the model of a non-stationary accretion disk.

We note note that a brief qualitative explanation of these lags in AGNs, very similar to the GBK13  quantitative model, was suggested by Marshall, Ryle \& Miller (\citeyear{2008ApJ...677..880M}). The conclusions of this model evidently contradict observational data. Therefore we consider another model for the formation of
the time delay in optical - X ray flashes from AGN.

\section{Model of a time lag formation in AGNs}

The accretion onto an SMBH in AGN takes place from surrounding gas, presumably formed by stellar winds of surrounding bulge stars. The angular momentum of stars in a quasi-spherical bulge is low, therefore formation of an accretion disk may
not occur, and accretion could take place in the form of a spherical flow.  The process of the tidal disruption of the star that approaches
an SMBH was investigated numerically in different approximations, see, for instance, Carter \& Luminet (\citeyear{1983A&A...121...97C}),
Frolov et al. (\citeyear{1994ApJ...432..680F}), Diener et al. (\citeyear{1995MNRAS.275..498D}), Marck, Lioure \& Bonazzola (\citeyear{1996A&A...306..666M}), Ivanov, Chernyakova, \& Novikov (\citeyear{2003MNRAS.338..147I}), and Ivanov \& Chernyakova (\citeyear{2006A&A...448..843I}).

  The flashes in AGNs, which are presumably connected to tidal disruptions of the surrounding stars in a close encounter with an SMBH, are accompanied by rapid ejection of matter with the
formation of a jet flowing outside, and another rapid jet directed toward the SMBH. A large part of the inner jet moves to the SMBH with a high velocity that is on the order of the free-fall velocity. The tidal disruption leads to the optical flash, and the X-ray flash starts when the matter of the inner jet is sufficiently heated to radiate in the X-ray region. It is not clear from the numerical simulations how the angular momentum is distributed over the matter falling onto the SMBH, nor which part of the matter falls with a small impact parameter onto the SMBH. We suggest that in the observed events this part is sufficiently large to create a strong initial flash.

The time delay between the optical and X-ray flashes follows from observations, listed above; the radius at which the optical flash occurs is calculated for the  motion with free-fall velocity $v_{ff}$ as

 \begin{equation}
v_{ff}=\sqrt{\frac{2GM}{r}},\quad \frac{dr}{dt}=v_{ff},\quad \tau_{ff}=\frac{2}{3}\frac{r^{3/2}}{\sqrt{GM}}.
\label{agn1}
\end{equation}
Taking $\tau_{ff}$ equal to observational time delay $\tau_{obs}$, we obtain a radius of the optical  flash $r_{opt}$ as

 \begin{equation}
r_{opt}=1.65\times 10^{12}\tau_{obs}m^{1/3}\,\,{\rm cm},
\label{agn2}
\end{equation}
where $\tau_{obs}$ is expressed in days and the SMBH mass $m$ is given in solar masses.

A tidal disruption of a star occurs when the tidal force from
the SMBH $F_t$ becomes comparable to the  gravitational force of the star $F_s$ at the radius of a star $R_s$ with a mass $M_s$.
The radius of the tidal disruption $r_t$ when these two forces become equal is written as, see, for example, Bisnovatyi-Kogan et al. (\citeyear{1982A&A...113..179B}),

 \begin{eqnarray}
F_t=2\frac{GM}{r^3}R_s,\quad F_s=\frac{GM_s}{R_s^2},  \nonumber\\
\quad r_t=R_s\left(2\frac{M_{BH}}{M_s}\right)^{1/3}=R_s(2\frac{m}{m_s})^{1/3}.
\label{agn3}
\end{eqnarray}

\begin{table}[h]
\caption{\label{t7} Properties of stars tidally disrupted by SMBH in AGN's.}
\centering
\begin{tabular}{lccc}
\hline\hline
Source&${\rm r_{opt}=r_t}$&${\rm R_S}$\\
\hline
\\
{\bf Mrk 509}           &$5.2\times 10^{15} {\rm cm}$    &$114\,m_s^{1/3} R_\odot $\\
{\bf NGC 7469}         &$8.95\times 10^{14} {\rm cm}$ &$47\,m_s^{1/3} R_\odot$\\
{\bf 3C 120} &$3.3\times 10^{15} {\rm cm}$&$100\,m_s^{1/3} R_\odot$\\
{\bf NGC 3516}       &$1.1\times 10^{16} {\rm cm}$ &$409\,m_s^{1/3} R_\odot$ \\
{\bf NGC 4051}     &$2.5\times 10^{14} {\rm cm}$&$36\,m_s^{1/3} R_\odot$\\
\hline
\end{tabular}
\tablefoot{%
Mass values of the disrupted stars $m_s=\frac{M_s}{M_\odot}$, the distance from the central black hole at which the optical flash occurs
$r_{opt}$, and radii of  stars
$R_s$ that are disrupted at the tidal radius $r_t$, identified with $r_{opt}=r_t$, are given for the SMBH in the AGNs listed above.
}
\end{table}
We see that the flashes in AGN that originate in tidal disruption may occur when a giant star with a radius of between a few tens and a few hundreds of solar radii enters the tidal radius. The observational time lag between the optical and X-ray flashes is quite consistent with the model we considered here, in which the optical flash occurs at the radius of the tidal disruption
and the X-ray flash occurs when the matter that is accreted with the free-fall velocity becomes hot enough as a result of adiabatic heating.

We note that another possibility to interpret the short time delays in AGNs is based on the irradiation model, see, for
example, Ulrich, Maraschi, and Urry  (\citeyear{1997ARA&A..35..445U}). This model could explain recent extensive observations of NGC 5548 in X-rays (SWIFT), UV, and optical light (HST) in the “reverberation mapping” campaign, see
Edelson et al. (\citeyear{2015ApJ...806..129E}) and Fausnaugh et al. (\citeyear{2016ApJ...821...56F}). In this object the UV and optical light curves are lagging the X-rays in short (one-
to two-day) intervals, so that the lag time increases for longer wavelengths. The irradiation model qualitatively explains this behavior quite well, but the estimated size of the disk is a factor of three larger than the prediction from standard thin-disk theory (Fausnaugh et al. \citeyear{2016ApJ...821...56F}). A sample of 21 active galactic nuclei was analyzed with data from the Swift
satellite to study the variability properties of the population in the X-ray, UV, and optical band (Buisson et al. \citeyear{2016arXiv160908638B}). A correlated variability between the emission in X-rays and UV is significantly detected for 9 of the 21 sources and is consistent with the UV lagging the X-rays. This behavior would be seen if the correlated UV variations were produced by the reprocessing of X-ray emission. In all cases the observed UV
lags are somewhat longer than expected for a standard thin disk. This may be connected to the incompleteness of the local accretion disk theory, see Novikov and Thorne (\citeyear{1973blho.conf..343N}).


\section{Conclusions}

We developed models of time lags between optical and X-ray flashes for close-binary galactic sources with accretion disks and for an AGN with an SMBH that is embedded in a quasi-spherical bulge.

    The time lag in disk-accreting galactic close-binary sources is based on a sudden increase in the accretion flow that starts at the disk periphery and is related to the optical maximum. The massive accretion layer propagates to the central compact source as a result of the turbulent viscosity. The X-ray flash occurs when this massive layer reaches the inner hot regions of the accretion disk and falls into the central compact object. The matter in the accretion disk  moves inside with a speed that is determined by the turbulent viscosity. We described this model quantitatively and derived an analytic formula that determines the value of the time lag. This formula gives results that agree well with observational values.

     The flashes in an AGN are considered in the model when a disruption  of a star that is in the evolution phase of a giant enters the radius of strong tidal forces. The matter with a low angular momentum that is released by the star  falls into the SMBH in the form of a quasi-spherical flow with a velocity
that is close to the free-fall velocity. An X-ray flash occurss when the falling matter reaches the hot inner regions. The time lag observed in these sources is identified with the time of the matter falling from the tidal radius onto the central region. The values of the tidal radius that we calculated in this model were compared with the theoretical radii of a tidal disruption
that depends on the masses of the SMBH and of the star, and on the radius of the star. Knowing the SMBH masses from observations, and making a reasonable suggestion for the stellar mass that
is on the order of one solar mass, we obtained that the radii of the disrupted star are between a few tens and a few hundreds of  $R_\odot$. These radii are characteristic of stars of moderate mass on the giant phase of evolution, see, for instance, Bisnovatyi-Kogan (2011).

    The matter with larger angular momentum that appeared in the disruption of the star is expected to form an accretion disk through which the matter will move to the center as a result
of turbulent viscosity, similarly to flashes in close galactic binaries. This motion is much slower than free-fall velocity and may last for many years. After such a flash in AGNs, we therefore expect a long-duration  irregular variability in the whole electromagnetic spectrum. As may be seen in Fig. 1 of Miller et al. (\citeyear{2015Natur.526..542M}), the X-ray and optical luminosity enter the non-steady plateau with a nonzero signal, which may be connected with a transition to the disk-accretion stage.

   The variability properties observed in many AGNs, where optical and UV emission lags the Xray light curve, may be explained by the model in which an X-ray flash in the center of AGN is followed by reradiation of the surrounding accretion disk.


\begin{acknowledgements}
     The work of GSBK was partially supported by the Russian Foundation for Basic Research Grant No. 14-02-00728 and No. OFI-M 14-29-06045,
 the Russian Federation President Grant for Support of Leading Scientific Schools, Grant No. NSh-261.2014.2, and  by Basic Research Program P-7 of the Presidium of the Russian Academy of Sciences.\\

We are indebted to NASAs {ADS}
for its magnificent literature and bibliography serving.
\end{acknowledgements}

\bibliographystyle{aa-note} 
\bibliography{timelag}    

\begin{thebibliography}{50}
\expandafter\ifx\csname natexlab\endcsname\relax\def\natexlab#1{#1}\fi

\bibitem[{{Alston} {et~al.}(2013){Alston}, {Vaughan}, \&
  {Uttley}}]{2013MNRAS.429...75A}
{Alston}, W.~N., {Vaughan}, S., \& {Uttley}, P. 2013, \mnras, 429, 75 \csname
  2013MNRAS.429...75Alink\endcsname~\csname 2013MNRAS.429...75Anote\endcsname

\bibitem[{{Bisnovatyi-Kogan}(2011)}]{2011spse.book.....B}
{Bisnovatyi-Kogan}, G.~S. 2011, {Stellar Physics 2: Stellar Evolution and
  Stability.} (Springer-Verlag Berlin-Heidelberg) \csname
  2011spse.book.....Blink\endcsname~\csname 2011spse.book.....Bnote\endcsname

\bibitem[{{Bisnovatyi-Kogan} {et~al.}(1982){Bisnovatyi-Kogan}, {Churaev}, \&
  {Kolosov}}]{1982A&A...113..179B}
{Bisnovatyi-Kogan}, G.~S., {Churaev}, R.~S., \& {Kolosov}, B.~I. 1982, \aap,
  113, 179 \csname 1982A&A...113..179Blink\endcsname~\csname
  1982A&A...113..179Bnote\endcsname

\bibitem[{{Bisnovatyi-Kogan} {et~al.}(2014){Bisnovatyi-Kogan}, {Klepnev}, \&
  {Giovannelli}}]{2014IJMPS..2860186B}
{Bisnovatyi-Kogan}, G.~S., {Klepnev}, A.~S., \& {Giovannelli}, F. 2014,
  International Journal of Modern Physics Conference Series, 28, 1460186
  \csname 2014IJMPS..2860186Blink\endcsname~\csname
  2014IJMPS..2860186Bnote\endcsname

\bibitem[{{Buisson} {et~al.}(2016){Buisson}, {Lohfink}, {Alston}, \&
  {Fabian}}]{2016arXiv160908638B}
{Buisson}, D.~J.~K., {Lohfink}, A.~M., {Alston}, W.~N., \& {Fabian}, A.~C.
  2016, ArXiv e-prints \csname 2016arXiv160908638Blink\endcsname~\csname
  2016arXiv160908638Bnote\endcsname

\bibitem[{{Carter} \& {Luminet}(1983)}]{1983A&A...121...97C}
{Carter}, B. \& {Luminet}, J.-P. 1983, \aap, 121, 97 \csname
  1983A&A...121...97Clink\endcsname~\csname 1983A&A...121...97Cnote\endcsname

\bibitem[{{De Loore} {et~al.}(1984){De Loore}, {Giovannelli}, {van Dessel},
  {Bartolini}, {Burger}, {Ferrari-Toniolo}, {Giangrande}, {Guarnieri},
  {Hellings}, {Hensberge}, {Persi}, {Piccioni}, \& {van
  Diest}}]{1984A&A...141..279D}
{De Loore}, C., {Giovannelli}, F., {van Dessel}, E.~L., {et~al.} 1984, \aap,
  141, 279 \csname 1984A&A...141..279Dlink\endcsname~\csname
  1984A&A...141..279Dnote\endcsname

\bibitem[{{de Martino} {et~al.}(1989){de Martino}, {Waters}, {Giovannelli}, \&
  {Persi}}]{1989ESASP.296..519D}
{de Martino}, D., {Waters}, L.~B.~F.~M., {Giovannelli}, F., \& {Persi}, P.
  1989, in ESA Special Publication, Vol. 296, Two Topics in X-Ray Astronomy,
  Volume 1: X Ray Binaries. Volume 2: AGN and the X Ray Background, ed.
  J.~{Hunt} \& B.~{Battrick} \csname 1989ESASP.296..519Dlink\endcsname~\csname
  1989ESASP.296..519Dnote\endcsname

\bibitem[{{Denney} {et~al.}(2010){Denney}, {Peterson}, {Pogge}, {Adair},
  {Atlee}, {Au-Yong}, {Bentz}, {Bird}, {Brokofsky}, {Chisholm}, {Comins},
  {Dietrich}, {Doroshenko}, {Eastman}, {Efimov}, {Ewald}, {Ferbey}, {Gaskell},
  {Hedrick}, {Jackson}, {Klimanov}, {Klimek}, {Kruse}, {Lad{\'e}route}, {Lamb},
  {Leighly}, {Minezaki}, {Nazarov}, {Onken}, {Petersen}, {Peterson},
  {Poindexter}, {Sakata}, {Schlesinger}, {Sergeev}, {Skolski}, {Stieglitz},
  {Tobin}, {Unterborn}, {Vestergaard}, {Watkins}, {Watson}, \&
  {Yoshii}}]{2010ApJ...721..715D}
{Denney}, K.~D., {Peterson}, B.~M., {Pogge}, R.~W., {et~al.} 2010, \apj, 721,
  715 \csname 2010ApJ...721..715Dlink\endcsname~\csname
  2010ApJ...721..715Dnote\endcsname

\bibitem[{{Diener} {et~al.}(1995){Diener}, {Kosovichev}, {Kotok}, {Novikov}, \&
  {Pethick}}]{1995MNRAS.275..498D}
{Diener}, P., {Kosovichev}, A.~G., {Kotok}, E.~V., {Novikov}, I.~D., \&
  {Pethick}, C.~J. 1995, \mnras, 275, 498 \csname
  1995MNRAS.275..498Dlink\endcsname~\csname 1995MNRAS.275..498Dnote\endcsname

\bibitem[{{Doroshenko} {et~al.}(2009){Doroshenko}, {Sergeev}, {Efimov},
  {Klimanov}, \& {Nazarov}}]{2009AstL...35..361D}
{Doroshenko}, V.~T., {Sergeev}, S.~G., {Efimov}, Y.~S., {Klimanov}, S.~A., \&
  {Nazarov}, S.~V. 2009, Astronomy Letters, 35, 361 \csname
  2009AstL...35..361Dlink\endcsname~\csname 2009AstL...35..361Dnote\endcsname

\bibitem[{{Edelson} {et~al.}(2015){Edelson}, {Gelbord}, {Horne}, {McHardy},
  {Peterson}, {Ar{\'e}valo}, {Breeveld}, {De Rosa}, {Evans}, {Goad}, {Kriss},
  {Brandt}, {Gehrels}, {Grupe}, {Kennea}, {Kochanek}, {Nousek}, {Papadakis},
  {Siegel}, {Starkey}, {Uttley}, {Vaughan}, {Young}, {Barth}, {Bentz},
  {Brewer}, {Crenshaw}, {Dalla Bont{\`a}}, {De Lorenzo-C{\'a}ceres}, {Denney},
  {Dietrich}, {Ely}, {Fausnaugh}, {Grier}, {Hall}, {Kaastra}, {Kelly},
  {Korista}, {Lira}, {Mathur}, {Netzer}, {Pancoast}, {Pei}, {Pogge},
  {Schimoia}, {Treu}, {Vestergaard}, {Villforth}, {Yan}, \&
  {Zu}}]{2015ApJ...806..129E}
{Edelson}, R., {Gelbord}, J.~M., {Horne}, K., {et~al.} 2015, \apj, 806, 129
  \csname 2015ApJ...806..129Elink\endcsname~\csname
  2015ApJ...806..129Enote\endcsname

\bibitem[{{Fausnaugh} {et~al.}(2016){Fausnaugh}, {Denney}, {Barth}, {Bentz},
  {Bottorff}, {Carini}, {Croxall}, {De Rosa}, {Goad}, {Horne}, {Joner},
  {Kaspi}, {Kim}, {Klimanov}, {Kochanek}, {Leonard}, {Netzer}, {Peterson},
  {Schn{\"u}lle}, {Sergeev}, {Vestergaard}, {Zheng}, {Zu}, {Anderson},
  {Ar{\'e}valo}, {Bazhaw}, {Borman}, {Boroson}, {Brandt}, {Breeveld}, {Brewer},
  {Cackett}, {Crenshaw}, {Dalla Bont{\`a}}, {De Lorenzo-C{\'a}ceres},
  {Dietrich}, {Edelson}, {Efimova}, {Ely}, {Evans}, {Filippenko}, {Flatland},
  {Gehrels}, {Geier}, {Gelbord}, {Gonzalez}, {Gorjian}, {Grier}, {Grupe},
  {Hall}, {Hicks}, {Horenstein}, {Hutchison}, {Im}, {Jensen}, {Jones},
  {Kaastra}, {Kelly}, {Kennea}, {Kim}, {Korista}, {Kriss}, {Lee}, {Lira},
  {MacInnis}, {Manne-Nicholas}, {Mathur}, {McHardy}, {Montouri}, {Musso},
  {Nazarov}, {Norris}, {Nousek}, {Okhmat}, {Pancoast}, {Papadakis}, {Parks},
  {Pei}, {Pogge}, {Pott}, {Rafter}, {Rix}, {Saylor}, {Schimoia}, {Siegel},
  {Spencer}, {Starkey}, {Sung}, {Teems}, {Treu}, {Turner}, {Uttley},
  {Villforth}, {Weiss}, {Woo}, {Yan}, \& {Young}}]{2016ApJ...821...56F}
{Fausnaugh}, M.~M., {Denney}, K.~D., {Barth}, A.~J., {et~al.} 2016, \apj, 821,
  56 \csname 2016ApJ...821...56Flink\endcsname~\csname
  2016ApJ...821...56Fnote\endcsname

\bibitem[{{Frolov} {et~al.}(1994){Frolov}, {Khokhlov}, {Novikov}, \&
  {Pethick}}]{1994ApJ...432..680F}
{Frolov}, V.~P., {Khokhlov}, A.~M., {Novikov}, I.~D., \& {Pethick}, C.~J. 1994,
  \apj, 432, 680 \csname 1994ApJ...432..680Flink\endcsname~\csname
  1994ApJ...432..680Fnote\endcsname

\bibitem[{{Giovannelli} {et~al.}(2007){Giovannelli}, {Bernabei}, {Rossi}, \&
  {Sabau-Graziati}}]{2007A&A...475..651G}
{Giovannelli}, F., {Bernabei}, S., {Rossi}, C., \& {Sabau-Graziati}, L. 2007,
  \aap, 475, 651 \csname 2007A&A...475..651Glink\endcsname~\csname
  2007A&A...475..651Gnote\endcsname

\bibitem[{{Giovannelli} {et~al.}(2015){Giovannelli}, {Bisnovatyi-Kogan},
  {Bruni}, {Corfini}, {Martinelli}, \& {Rossi}}]{2015AcA....65..107G}
{Giovannelli}, F., {Bisnovatyi-Kogan}, G.~S., {Bruni}, I., {et~al.} 2015,
  \actaa, 65, 107 \csname 2015AcA....65..107Glink\endcsname~\csname
  2015AcA....65..107Gnote\endcsname

\bibitem[{{Giovannelli} {et~al.}(2013){Giovannelli}, {Bisnovatyi-Kogan}, \&
  {Klepnev}}]{2013A&A...560A...1G}
{Giovannelli}, F., {Bisnovatyi-Kogan}, G.~S., \& {Klepnev}, A.~S. 2013, \aap,
  560, A1 (GBK13) \csname 2013A&A...560A...1Glink\endcsname~\csname
  2013A&A...560A...1Gnote\endcsname

\bibitem[{{Giovannelli} {et~al.}(1983){Giovannelli}, {Gaudenzi}, {Rossi}, \&
  {Piccioni}}]{1983AcA....33..319G}
{Giovannelli}, F., {Gaudenzi}, S., {Rossi}, C., \& {Piccioni}, A. 1983, \actaa,
  33, 319 \csname 1983AcA....33..319Glink\endcsname~\csname
  1983AcA....33..319Gnote\endcsname

\bibitem[{{Giovannelli} \& {Sabau-Graziati}(2011)}]{2011AcPol..51b..21G}
{Giovannelli}, F. \& {Sabau-Graziati}, L. 2011, Acta Polytechnica, 51, 21
  \csname 2011AcPol..51b..21Glink\endcsname~\csname
  2011AcPol..51b..21Gnote\endcsname

\bibitem[{{Giovannelli} \& {Sabau-Graziati}(2012)}]{2012AcPol..52a..11G}
{Giovannelli}, F. \& {Sabau-Graziati}, L. 2012, Acta Polytechnica, 52, 11
  \csname 2012AcPol..52a..11Glink\endcsname~\csname
  2012AcPol..52a..11Gnote\endcsname

\bibitem[{{Hameury} {et~al.}(1999){Hameury}, {Lasota}, \&
  {Dubus}}]{1999MNRAS.303...39H}
{Hameury}, J.-M., {Lasota}, J.-P., \& {Dubus}, G. 1999, \mnras, 303, 39 \csname
  1999MNRAS.303...39Hlink\endcsname~\csname 1999MNRAS.303...39Hnote\endcsname

\bibitem[{{Ivanov} \& {Chernyakova}(2006)}]{2006A&A...448..843I}
{Ivanov}, P.~B. \& {Chernyakova}, M.~A. 2006, \aap, 448, 843 \csname
  2006A&A...448..843Ilink\endcsname~\csname 2006A&A...448..843Inote\endcsname

\bibitem[{{Ivanov} {et~al.}(2003){Ivanov}, {Chernyakova}, \&
  {Novikov}}]{2003MNRAS.338..147I}
{Ivanov}, P.~B., {Chernyakova}, M.~A., \& {Novikov}, I.~D. 2003, \mnras, 338,
  147 \csname 2003MNRAS.338..147Ilink\endcsname~\csname
  2003MNRAS.338..147Inote\endcsname

\bibitem[{{Karakula} {et~al.}(1984){Karakula}, {Tkaczyk}, \&
  {Giovannelli}}]{1984AdSpR...3..335K}
{Karakula}, S., {Tkaczyk}, W., \& {Giovannelli}, F. 1984, Advances in Space
  Research, 3, 335 \csname 1984AdSpR...3..335Klink\endcsname~\csname
  1984AdSpR...3..335Knote\endcsname

\bibitem[{{Landau} \& {Lifshitz}(1980)}]{1980stph.book.....L}
{Landau}, L.~D. \& {Lifshitz}, E.~M. 1980, {Statistical physics. Pt.1} (Oxford:
  Pergamon Press) \csname 1980stph.book.....Llink\endcsname~\csname
  1980stph.book.....Lnote\endcsname

\bibitem[{{Lasota}(2001)}]{2001NewAR..45..449L}
{Lasota}, J.-P. 2001, \nar, 45, 449 \csname
  2001NewAR..45..449Llink\endcsname~\csname 2001NewAR..45..449Lnote\endcsname

\bibitem[{{Maoz} {et~al.}(2000){Maoz}, {Edelson}, \&
  {Nandra}}]{2000AJ....119..119M}
{Maoz}, D., {Edelson}, R., \& {Nandra}, K. 2000, \aj, 119, 119 \csname
  2000AJ....119..119Mlink\endcsname~\csname 2000AJ....119..119Mnote\endcsname

\bibitem[{{Marck} {et~al.}(1996){Marck}, {Lioure}, \&
  {Bonazzola}}]{1996A&A...306..666M}
{Marck}, J.~A., {Lioure}, A., \& {Bonazzola}, S. 1996, \aap, 306, 666 \csname
  1996A&A...306..666Mlink\endcsname~\csname 1996A&A...306..666Mnote\endcsname

\bibitem[{{Marshall} {et~al.}(2008){Marshall}, {Ryle}, \&
  {Miller}}]{2008ApJ...677..880M}
{Marshall}, K., {Ryle}, W.~T., \& {Miller}, H.~R. 2008, \apj, 677, 880 \csname
  2008ApJ...677..880Mlink\endcsname~\csname 2008ApJ...677..880Mnote\endcsname

\bibitem[{{Mehdipour} {et~al.}(2011){Mehdipour}, {Branduardi-Raymont},
  {Kaastra}, {Petrucci}, {Kriss}, {Ponti}, {Blustin}, {Paltani}, {Cappi},
  {Detmers}, \& {Steenbrugge}}]{2011A&A...534A..39M}
{Mehdipour}, M., {Branduardi-Raymont}, G., {Kaastra}, J.~S., {et~al.} 2011,
  \aap, 534, A39 \csname 2011A&A...534A..39Mlink\endcsname~\csname
  2011A&A...534A..39Mnote\endcsname

\bibitem[{{Meshcheryakov} {et~al.}(2013){Meshcheryakov}, {Khamitov},
  {Eselevich}, {Pavlinsky}, {Burenin}, {Bikmaev}, {Melnikov}, {Galeev},
  {Kirbiyik}, {Uluc}, {Kaynar}, \& {Okuyan}}]{2013ATel.5114....1M}
{Meshcheryakov}, A., {Khamitov}, I., {Eselevich}, M., {et~al.} 2013, The
  Astronomer's Telegram, 5114 \csname 2013ATel.5114....1Mlink\endcsname~\csname
  2013ATel.5114....1Mnote\endcsname

\bibitem[{{Miller} {et~al.}(2015){Miller}, {Kaastra}, {Miller}, {Reynolds},
  {Brown}, {Cenko}, {Drake}, {Gezari}, {Guillochon}, {Gultekin}, {Irwin},
  {Levan}, {Maitra}, {Maksym}, {Mushotzky}, {O'Brien}, {Paerels}, {de Plaa},
  {Ramirez-Ruiz}, {Strohmayer}, \& {Tanvir}}]{2015Natur.526..542M}
{Miller}, J.~M., {Kaastra}, J.~S., {Miller}, M.~C., {et~al.} 2015, \nat, 526,
  542 \csname 2015Natur.526..542Mlink\endcsname~\csname
  2015Natur.526..542Mnote\endcsname

\bibitem[{{Nagase} {et~al.}(1982){Nagase}, {Hayakawa}, {Kunieda}, {Makino},
  {Masai}, {Tawara}, {Inoue}, {Kawai}, {Koyama}, {Makishima}, {Matsuoka},
  {Murakami}, {Oda}, {Ogawara}, {Ohashi}, {Shibazaki}, {Tanaka}, {Miyamoto},
  {Tsunemi}, {Yamashita}, \& {Kondo}}]{1982ApJ...263..814N}
{Nagase}, F., {Hayakawa}, S., {Kunieda}, H., {et~al.} 1982, \apj, 263, 814
  \csname 1982ApJ...263..814Nlink\endcsname~\csname
  1982ApJ...263..814Nnote\endcsname

\bibitem[{{Nandra} {et~al.}(1998){Nandra}, {Clavel}, {Edelson}, {George},
  {Malkan}, {Mushotzky}, {Peterson}, \& {Turner}}]{1998ApJ...505..594N}
{Nandra}, K., {Clavel}, J., {Edelson}, R.~A., {et~al.} 1998, \apj, 505, 594
  \csname 1998ApJ...505..594Nlink\endcsname~\csname
  1998ApJ...505..594Nnote\endcsname

\bibitem[{{Novikov} \& {Thorne}(1973)}]{1973blho.conf..343N}
{Novikov}, I.~D. \& {Thorne}, K.~S. 1973, in Black Holes (Les Astres Occlus),
  ed. C.~{Dewitt} \& B.~S. {Dewitt}, 343--450 \csname
  1973blho.conf..343Nlink\endcsname~\csname 1973blho.conf..343Nnote\endcsname

\bibitem[{{Onken} {et~al.}(2004){Onken}, {Ferrarese}, {Merritt}, {Peterson},
  {Pogge}, {Vestergaard}, \& {Wandel}}]{2004ApJ...615..645O}
{Onken}, C.~A., {Ferrarese}, L., {Merritt}, D., {et~al.} 2004, \apj, 615, 645
  \csname 2004ApJ...615..645Olink\endcsname~\csname
  2004ApJ...615..645Onote\endcsname

\bibitem[{{Orosz} \& {Bailyn}(1997)}]{1997ApJ...477..876O}
{Orosz}, J.~A. \& {Bailyn}, C.~D. 1997, \apj, 477, 876 \csname
  1997ApJ...477..876Olink\endcsname~\csname 1997ApJ...477..876Onote\endcsname

\bibitem[{{Orosz} {et~al.}(1997){Orosz}, {Remillard}, {Bailyn}, \&
  {McClintock}}]{1997ApJ...478L..83O}
{Orosz}, J.~A., {Remillard}, R.~A., {Bailyn}, C.~D., \& {McClintock}, J.~E.
  1997, \apjl, 478, L83 \csname 1997ApJ...478L..83Olink\endcsname~\csname
  1997ApJ...478L..83Onote\endcsname

\bibitem[{{Peterson} {et~al.}(2004){Peterson}, {Ferrarese}, {Gilbert}, {Kaspi},
  {Malkan}, {Maoz}, {Merritt}, {Netzer}, {Onken}, {Pogge}, {Vestergaard}, \&
  {Wandel}}]{2004ApJ...613..682P}
{Peterson}, B.~M., {Ferrarese}, L., {Gilbert}, K.~M., {et~al.} 2004, \apj, 613,
  682 \csname 2004ApJ...613..682Plink\endcsname~\csname
  2004ApJ...613..682Pnote\endcsname

\bibitem[{{Shahbaz} {et~al.}(1998){Shahbaz}, {Bandyopadhyay}, {Charles},
  {Wagner}, {Muhli}, {Hakala}, {Casares}, \& {Greenhill}}]{1998MNRAS.300.1035S}
{Shahbaz}, T., {Bandyopadhyay}, R.~M., {Charles}, P.~A., {et~al.} 1998, \mnras,
  300, 1035 \csname 1998MNRAS.300.1035Slink\endcsname~\csname
  1998MNRAS.300.1035Snote\endcsname

\bibitem[{{Shakura}(1972)}]{1972AZh....49..921S}
{Shakura}, N.~I. 1972, \azh, 49, 921 \csname
  1972AZh....49..921Slink\endcsname~\csname 1972AZh....49..921Snote\endcsname

\bibitem[{{Shemmer} {et~al.}(2003){Shemmer}, {Uttley}, {Netzer}, \&
  {McHardy}}]{2003MNRAS.343.1341S}
{Shemmer}, O., {Uttley}, P., {Netzer}, H., \& {McHardy}, I.~M. 2003, \mnras,
  343, 1341 \csname 2003MNRAS.343.1341Slink\endcsname~\csname
  2003MNRAS.343.1341Snote\endcsname

\bibitem[{{Smak}(1984)}]{1984AcA....34..161S}
{Smak}, J. 1984, \actaa, 34, 161 \csname
  1984AcA....34..161Slink\endcsname~\csname 1984AcA....34..161Snote\endcsname

\bibitem[{{Smak}(1998)}]{1998AcA....48..677S}
{Smak}, J.~I. 1998, \actaa, 48, 677 \csname
  1998AcA....48..677Slink\endcsname~\csname 1998AcA....48..677Snote\endcsname

\bibitem[{{Ulrich} {et~al.}(1997){Ulrich}, {Maraschi}, \&
  {Urry}}]{1997ARA&A..35..445U}
{Ulrich}, M.-H., {Maraschi}, L., \& {Urry}, C.~M. 1997, \araa, 35, 445 \csname
  1997ARA&A..35..445Ulink\endcsname~\csname 1997ARA&A..35..445Unote\endcsname

\bibitem[{{Vestergaard} \& {Peterson}(2006)}]{2006ApJ...641..689V}
{Vestergaard}, M. \& {Peterson}, B.~M. 2006, \apj, 641, 689 \csname
  2006ApJ...641..689Vlink\endcsname~\csname 2006ApJ...641..689Vnote\endcsname

\bibitem[{{Voges} {et~al.}(1999){Voges}, {Aschenbach}, {Boller},
  {Br{\"a}uninger}, {Briel}, {Burkert}, {Dennerl}, {Englhauser}, {Gruber},
  {Haberl}, {Hartner}, {Hasinger}, {K{\"u}rster}, {Pfeffermann}, {Pietsch},
  {Predehl}, {Rosso}, {Schmitt}, {Tr{\"u}mper}, \&
  {Zimmermann}}]{1999A&A...349..389V}
{Voges}, W., {Aschenbach}, B., {Boller}, T., {et~al.} 1999, \aap, 349, 389
  \csname 1999A&A...349..389Vlink\endcsname~\csname
  1999A&A...349..389Vnote\endcsname

\bibitem[{{Waterhouse} {et~al.}(2016){Waterhouse}, {Degenaar}, {Wijnands},
  {Brown}, {Miller}, {Altamirano}, \& {Linares}}]{2016MNRAS.456.4001W}
{Waterhouse}, A.~C., {Degenaar}, N., {Wijnands}, R., {et~al.} 2016, \mnras,
  456, 4001 \csname 2016MNRAS.456.4001Wlink\endcsname~\csname
  2016MNRAS.456.4001Wnote\endcsname

\bibitem[{{Wheatley} {et~al.}(2003){Wheatley}, {Mauche}, \&
  {Mattei}}]{2003MNRAS.345...49W}
{Wheatley}, P.~J., {Mauche}, C.~W., \& {Mattei}, J.~A. 2003, \mnras, 345, 49
  \csname 2003MNRAS.345...49Wlink\endcsname~\csname
  2003MNRAS.345...49Wnote\endcsname

\bibitem[{{Yamaoka} {et~al.}(2011){Yamaoka}, {Krimm}, {Tomsick}, {Kaaret},
  {Kalemci}, {Corbel}, \& {Migliari}}]{2011ATel.3686....1Y}
{Yamaoka}, K., {Krimm}, H.~A., {Tomsick}, J.~A., {et~al.} 2011, The
  Astronomer's Telegram, 3686 \csname 2011ATel.3686....1Ylink\endcsname~\csname
  2011ATel.3686....1Ynote\endcsname

\end{thebibliography}
\end{document}